\documentstyle[psfig,jkas99]{article}
\runningauthor{MOON ET EL.}
\beginpage{1}
\endpage{8}
\begin{document}


\title{STUDY OF SOLAR ACTIVE REGIONS BASED ON BOAO VECTOR MAGNETOGRAMS}
\author{\large{\bf{YONG-JAE MOON$^{1}$, YOUNG DEUK PARK$^1$, HONG SIK YUN$^2$, \\AND EUN-AH CHO$^{1,3}$}}}
\address{\normalsize{$^1$Bohyunsan Optical 
Astronomical Observatory, Korea Astronomy
Observatory, Kyungpook 770-820}}
\address{\normalsize{$^2$Department of Astronomy, Seoul National University, Seoul 151-742}}
\address{\normalsize{$^3$Department of Astronomy and Atmospheric Sciences, 
Kyungpook National University, Kyungpook 702-701}}

\abstract{
In this study we present the study of solar active regions based on BOAO
vector magnetograms and $H\alpha$ filtergrams.
With  the new calibration method we analyzed BOAO vector
magnetograms taken from the SOFT observational system to compare with those of other observing systems. In this study it has been demonstrated  that
(1) our longitudinal magnetogram  matches very well the corresponding Mitaka's
 magnetogram to the extent that 
the maximum correlation yields r=0.962
  between our re-scaled longitudinal magnetogram
and the Mitaka's
 magnetogram; 
(2) according to a comparison of
our magnetograms of
AR 8422 with those taken at Mitaka 
 solar observatory their
longitudinal fields are very similar to each other
while transverse fields are a little different possibly due to large noise 
level; 
(3) main features seen by our longitudinal magnetograms
of AR 8422 and AR 8419 and the  corresponding Kitt Peak 
 magnetograms 
are very similar to each other;
(4)  time series of our vector magnetograms and H-alpha observations of  
 AR 8419 during its flaring (M3.1/1B) activity
 show that the filament eruption followed the sheared inversion line of the quadrupolar configuration of sunspots, indicating that the flare should be associated with
the quadrupolar field configuration and  
its interaction with new filament eruption.
Finally, it may be concluded that the Solar Flare Telescope at BOAO works normally and it is ready to do numerous observational and theoretical works associated with solar activities such as flares. }

\keywords{Sun: magnetic fields $-$ Sun : vector magnetogram $-$ Sun : flare}

\maketitle

\section{INTRODUCTION}

It is well known that magnetic fields play an important role
in solar active phenomena such as solar flares and prominences.
However, measurements of magnetic fields are 
only available at the photosphere
and very limitedly at the chromosphere.
Thus  the evolution of magnetic fields at the photosphere
has widely used for studies of  relationships between active phenomena and magnetic fields.
In this sense, reasonable measurements of solar magnetic fields at the
photospheric level are of key
importance in understanding solar activities
(e.g. Hagyard et al. 1984).

Solar Flare Telescope (SOFT) has been set up
at the peak of Mt. Bohyun in 1995.
A filter-based magnetograph (Vector Magnetograph, VMG)
 is attached to 
 the SOFT (Moon 1999c, Park et al. 1997) of Bohyunsan Optical Astronomy Observatory (BOAO), which
uses a very narrow band Lyot (birefringent) filter which  
 measures
magnetic fields at the solar photosphere with Fe I 6302.5 line. 
The Stokes parameters are measured by collecting spectrally
 integrated data  over the filter passband. It has very high time resolution  
 which is 
less than 1 minute, with 
 relatively  large
 field of view ($400'' \times 300''$).  
For the efficient use of the SOFT, we have developed
the data acquisition system (Moon et al. 1996), the telescope control software (Moon et al. 1997), the KD*P control system (Nam et al. 1997), and the four channel filter control system (Jang et al. 1998).

The calibration problem of filter-based magnetographs with Fe I 6302.5 spectral line
has been discussed by several authors (Ichimoto 1993,  Sakurai et al. 1995, Kim 1997, Moon et al. 199b). Recently, Moon et al. (1999b) have developed an improved calibration method  by using theoretical Stokes polarization signals calculated with
various inclination angles of magnetic fields (Hagyard and Kineke 1995).

In this paper   we  study solar active regions using BOAO vector magnetograms with the new calibration method.  For this we describe 
how to analyze BOAO vector magnetograms in Section II and compare observed magnetograms
with corresponding magnetograms from other solar observatories in Section III.
In Section IV we  present some observational results of AR 8419 by the
SOFT. A brief summary and conclusion will be given in Section V.

\section{ANALYSIS OF VECTOR MAGNETOGRAMS}

\subsection{Dark Frame and Flat Field Correction}
For detecting observed images through VMG we use a SONY-XC 77 video CCD,
whose signal is
digitized by the image processor (Moon et al. 1996).
Dark frame observations are made
by closing the cover of the telescope. 
Flat field observations were made
by  calibration optics to produce defocused lights. However, 
the observed flat images had relatively large
intensity gradients so that we now use a
defocused intensity image of the solar disk center
as a flat image by adjusting a focusing motor.

The dark frame (D) and flat field (F) corrections are made by
\begin{equation}
I_c(+) ={I_o(+) - D \over F(+) - D},
\end{equation}
\begin{equation}
I_c(-) ={I_o(-) - D \over F(-) - D},
\end{equation}
where $I_o$ represents an observed image  and $I_c$, the
image corrected for dark frame and flat field.
We found that there are little systematic difference between $F(+)$ and $F(-)$.
In Stokes V/I observation, 
a final image corrected for dark frame and flat field
can be expressed as
\begin{equation}
{V \over I} = { I_c(+) - I_c(-) \over I_c(+) + I_c(-)} =
  { I_o(+) - I_o(-) \over I_o(+) + I_o(-) - 2D},
\end{equation}
where we assume that $F(+)=F(-)$.  This argument can be also
applied to Stokes Q/I and U/I observations. These facts imply that
flat field correction  can be neglected in
 magnetic field observations.

 \subsection{Alignment of Q, U, V Images}

Due to atmospheric seeing and tracking instability,
observed images could be shifted during filter turret transition for
Q, U, and V observations.
When we make an observation of a  set of Stokes data,
three FITS files for Q, U and V are obtained, in which 
$I(+)$ and $I(-)$  are separately saved. 
Here $I(+)$ and  $I(-)$  correspond to a  filter 
convolved monochromatic intensity of right and left hand
polarization, respectively.
For convenience we define three intensity images as follows : 
$I_q = I_q(+)+I_q(-)$ in Q data, 
$I_u = I_u(+)+I_u(-)$ in U data, and
$I_v = I_v(+)+I_v(-)$ in V data. 
We usually make  alignments of theses intensity images by
shifting $I_q$ and $I_u$ images to match an $I_v$ image
as follows.
(1) Two set of coordinates for the same 
  largest sunspots in both a target  image ($I_u$ or $I_q$)  and
  a reference  image ($I_v$) are searched by 
 the center of gravity method (Ichimoto 1993).
(2) An appropriate size of window  around the coordinates are determined.
(3) The target image  is shifted to have a 
  maximum correlation
  with the reference  image for the selected window.

\subsection{Calibration}

Moon et al. (1999b) have already discussed the calibration problems of filter-based magnetograms,
especially focusing on the Fe I 6302.5 spectral line for the SOFT.
In applying  our developed method  to the actual analysis
the  following facts should be kept in mind.
In many cases, 
 magnetic field strengths
 derived from filter-based magnetographs  
have been underestimated relative to theoretically predicted
or spectrally determined ones. To compensate this
problem, an arbitrary factor, so called k-factor, 
has been introduced to raise the
observed  polarization  signals  
so   that  it  matches  the  field  strength   estimated  
from 
nonfilter-based magnetic observations (Gary et al. 1987,
 Chae 1996).
According to Chae (1996), stray light
 corrected
fields still require  a k-factor to match empirically determined ones. 
The underestimation of magnetic fields
might be due to stray light effect (Chae et al. 1998a, 1998b),
instrumental depolarization (Gary 1991), the fragmental distribution of the magnetic field on the solar surface (Ichimoto 1993), and
transmission wavelength error etc. 
To properly correct all the problems related to this matter
seems to be too challenging.

For the calibration of BOAO magnetograms, we suggest to select
one of two methods. The first method is a calibration method for Mitaka Solar Observatory (Ichimoto 1993, 1997, Sakurai et al. 1995), which is applicable
to BOAO vector magnetograms since two observational systems are very
similar to each other. 
Here we summarize Mitaka's method as follows.
(1) Observed polarization signals are converted
 to magnetic field strengths by the
method described in Ichimoto (1993).
(2) A k-factor is multiplied to
 longitudinal fields to balance 
between observed transverse fields
 and corresponding potential fields 
derived from the observed longitudinal fields (Sakurai et al. 1995). 
First, potential magnetic fields are
derived  by using a Fourier expansion method (Sakurai 1992) in which observed
longitudinal fields are used as a boundary condition. Then  they select data points on which the observed and the computed transverse fields have the nearly same directions, and compute the average ratio of the observed and computed 
transverse field strengths. Finally, this ratio (k-factor)
is multiplied to the observed
longitudinal fields to produce  re-scaled longitudinal ones.
In addition, we re-scale  the calibrated vector fields to balance
between the maximum of longitudinal fields and that of corresponding fields 
from a full disk longitudinal 
magnetogram of Kitt Peak Solar Observatory, which have
provided us with 
unique daily full disk longitudinal magnetograms by the NSO/KP
Vacuum Telescope together with a 10.7m vertical Littrow spectrograph
(Jones et al. 1992). 

In terms of
sunspot models as well as observational data of active regions,
we have tested the validity of the second process that
transverse fields  balance with corresponding potential fields 
derived from
longitudinal fields. 
First, we computed k-factors for three sunspot models which well
describe observed field configuration at the photosphere 
and  
computed values  are found to be 
1.12 for Skumanich (1992)'s  dipole
model, 1.09 for Yun(1970)'s sunspot model, and 1.05 for Moon et al. (1998)'s
sunspot model.
We have also computed k-factors for 37 vector 
magnetograms of four flare-productive  
active regions (AR 5747, AR 6233, AR 6659, and AR 6982) observed at Mees Solar
Observatory 
whose Stokes polarimeter is one of 
qualified spectrometer-based magnetographs.
It is found
 that computed values 
for 28 magnetograms
(about 76\%) out of 37 are approximately unity within 10\% accuracy.
These results imply that  the second process could be applied to 
representative active regions.
 
The second calibration method is to employ
the iterative calibration method (Moon et al. 1999b), which
 only  works as long as
both circular and linear polarization signals are reasonably estimated.
Unfortunately, underestimation of circular  polarization
signals are larger than that of linear polarization ones.
Thus we have devised an iterative method for calibration as follows. 
(1) We follow two-steps (1) and (2) of the first calibration
 method 
and multiply a derived k-factor to circular polarization signals.
(2) We convert both re-scaled circular 
and original linear polarization signals to vector fields
by our developed calibration method described in Moon et al. (1999b). 
(3) We re-scale longitudinal fields by the step (2) of the
first method.
(4) We iterate steps (2) and (3) until a k-factor (re-scaling factor)
  converges unity within 5 \% accuracy.
(5) If necessary, we re-scale again the calibrated vector fields
 to balance
between the maximum of longitudinal fields and that of corresponding fields 
from a full disk longitudinal 
magnetogram of Kitt Peak Solar Observatory.

\subsection{Solutions for $180^o$ ambiguity}

When one analyzes vector magnetograms
from solar magnetograph measurements, one of  challenging 
problems is  to solve
the $180^o$ ambiguity  in the azimuth of
observed transverse fields. This ambiguity
is attributed to the fact that two anti-parallel
polarization
signals of transverse fields are indistinguishable 
each other since the transverse measurements
of the magnetograph provides only the plane of
linear polarization.
For all vector magnetograms the ambiguity should be resolved
to obtain the correct transverse field components.
The great importance of the problem  is 
attributed to the fact
that the reasonable resolution of
the problem can give us
a meaningful understanding on
physical quantities such as vertical
current density, shear angle, magnetic free
energy.

To resolve the ambiguity, an additional
constraint on the field azimuth should be
introduced in terms of theoretical or observational aspects.
One of commonly used ways is the potential
field method based on the
fact that an observed transverse
magnetic field is not far 
away from a corresponding potential component. That is,
the direction of the transverse field is chosen
such that the two transverse components make an acute angle.
This method holds for         
for nearly
potential regions but not
for highly non-potential regions.

For our study we adopt two ambiguity methods :
a  potential field method (Sakurai 1992) and  a multi-step method
(Canfield et al. 1993). 
Comprehensive reviews  for resolving the problem
are found in several literatures
(e.g., Sakurai 1989, Wang 1993, Gary and Demoulin 1995). 

\subsubsection{Potential Field Method} 
For the case of a potential field, the magnetic field
can be derived from a scalar potential $\Phi$,
\begin{equation}
 {\bf B} = -\nabla  \Phi
\end{equation}
Using $\nabla \cdot {\bf B}=0$, the potential should
satisfy the Laplace's equation:
\begin{equation}
   \nabla^2 \Phi = 0,
\end{equation}
where an observable quantity $B_z$ is used as the
boundary condition, $ B_z = \partial \Phi / 
\partial z$. 

The potential field solution was
initially suggested by Schmidt (1964) with the
use of the Green function method. Later, 
the Fourier expansion method has been
 employed by Teuber et al.(1977) and
Sakurai(1989). For our study, we have used
a Fourier expansion method developed by
Sakurai (1992).
 The criterion 
of the method is given by
\begin{equation}
{\bf B_{ot}} \cdot {\bf B_{pt}} > 0,
\end{equation}
where ${\bf B_{ot}}$ is an observed
transverse field and ${\bf B_{pt}}$ is a tangential
component of the potential field
solution derived from  Equation (5).

\subsubsection{Multi-step Method}
Canfield et al. (1993) 
employed a multi-step for $180^o$ ambiguity solution,
which was well described in the Appendix of their paper.
Here we summarize their method as follows.
Step 1 : They first choose the orientation of each transverse field vector which is closest to the potential field computed using longitudinal fields
as a boundary condition. Then they rotate the data to the heliographic coordinate system.
Step 2 : For current-carrying active regions, after resolving with the potential field, they compute the linear force-free field with a linear force-free coefficient $\alpha$ selected to match
nonpotentialities discovered in Step 1.
Step 3 : They next choose the orientation of the transverse field 
which minimizes the angle
between neighboring field vectors by maximizing the sum of
the vector dot product of the field vector with each of its eight neighbors.
Step 4 :  In regions with strong total magnetic field strength ($\ge 1000G$) and a high degree of shear
(transverse field azimuth differing from the potential azimuth by more than $85^o$), they iteratively select the orientation of the field which minimizes
the field divergence 
$|\nabla \cdot {\bf B}|$.
Step 5 : Finally, in regions where the total field strength is below the noise level in the magnetograms, they iteratively choose the orientation of the field which
minimizes the electric current.

\section{COMPARISON WITH OTHER MAGNETOGRAMS}
 
We have 
compared vector fields of AR 8422 with
corresponding magnetogram of Mitaka Solar Telescope (Sakurai et al.
1995) which have
produced qualified vector magnetograms since 1992.
The comparison seems to be more meaningful in
that instruments and detecting system
of the Mitaka Solar Observatory are very similar
 to those of the SOFT in BOAO. 

 For comparison
 we adopted the first calibration method, i.e., 
the standard reduction
procedure of Mitaka Observatory (Ichimoto 1993, 1997), and 
the multi-step method (Canfield et al. 1993) for the
$180^o$ ambiguity resolution.
 
A field of view ( $400''\times300''$)
 of VMG in the SOFT was originally
 determined from its optical layout.
Since the optical layout have been a little changed  
so that its field of view
needs to be redetermined.
In the case of Mitaka's magnetograms,
its field of view was determined by using
a stop tracker (Ichimoto 1993).
We  have determined a field of view
 of vector magnetogram
made with VMG by comparing with Mitaka's 
corresponding magnetograms as follows.
First of all, we have made a linear matching of SOFT's magnetogram
with Mitaka's  corresponding magnetogram for AR 8422
(S23W38) on Dec. 28, 1998.
For this we make a linear mapping
 of reference magnetogram over the image
coordinate system of corresponding 
magnetogram under consideration (Chae 1999):
\begin{equation}
i = S_x l + x_0,
\end{equation}
\begin{equation}
j= S_y m + y_0,
\end{equation}
where $l$ and $m$ are coordinates of data points
 in a considered magnetogram in unit of pixel, and
 $i$  and  $j$  are  coordinates of 
 the  corresponding  points  in  a  reference 
magnetogram.  Chae (1999) determined four parameters by minimizing
\begin{equation}
H= \sum ( C_{lm} - R_{ij} )^2
\end{equation}
over $S_x$, $S_y$, $x_0$, and $y_0$, where $C_{lm}$ is a 
target magnetogram and $R_{ij}$ is a re-mapped reference magnetogram.

We applied Chae's method to a longitudinal magnetogram of AR 8422
observed on Dec. 28, 1998
and corresponding Mitaka's magnetogram.
We have also developed another method which  
derive i, j, l, and m  to have a maximum correlation
between $C_{lm}$ and $R_{ij}$.
Two methods are in good agreements with each other, as expected.
The field of view ($400''\times300''$)
of our observed vector magnetogram
 has been confirmed  by finding
the maximum correlation between our remapped longitudinal magnetogram
and corresponding Mitaka's
 magnetogram within about 1\%.
Figure 1 shows
the comparison of
BOAO's longitudinal magnetic fields of
AR 8422 and corresponding magnetic fields made with a
similar magnetograph at Mitaka Solar Observatory.
 As seen in the figure,
they are well correlated with each other (r=0.962).
Some differences in negative strong field regions 
may be due to seeing condition,
tracking instability, filter transmission wavelength, and observational time difference etc.

Figure 2 shows BOAO(upper) and Mitaka(lower) vector magnetic fields of AR 8422
observed on Dec. 28, 1998. Its main features of longitudinal fields
are very similar to each other, while those of transverse fields are 
a little
different, which might be due to large noise levels ( $100 \sim 200$ G) of
transverse fields.

\begin{figure}[t]
\psfig{figure=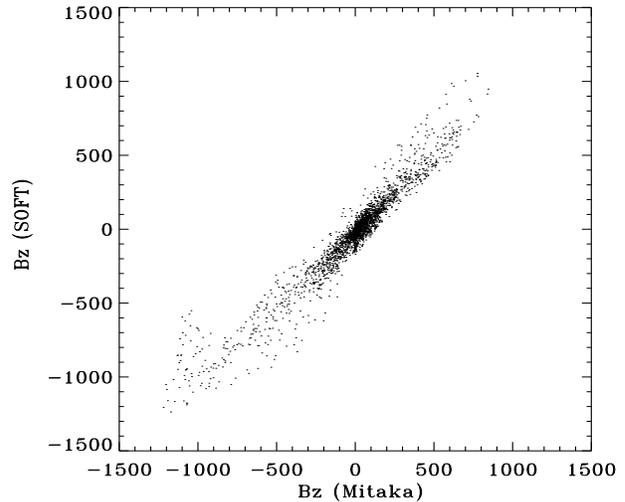,height=7cm,width=8cm}
\caption{
Comparison of two longitudinal magnetic field components of
AR 8422 acquired by
BOAO and Mitaka's magnetographs.
These two longitudinal fields  are well correlated with
each other (r=0.962).
}
\end{figure}

\begin{figure}
\psfig{figure=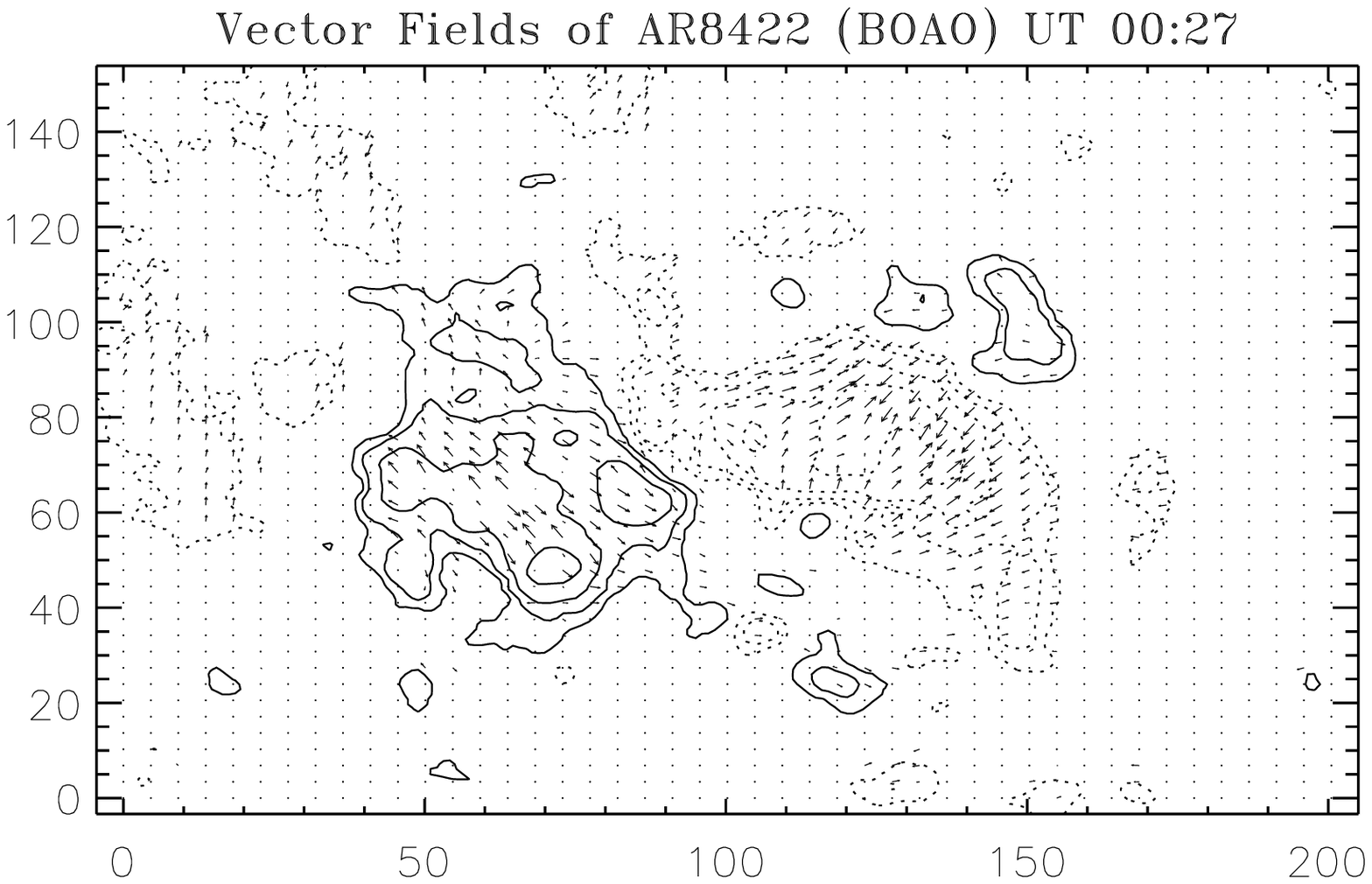,height=7.2cm,width=8cm}
\psfig{figure=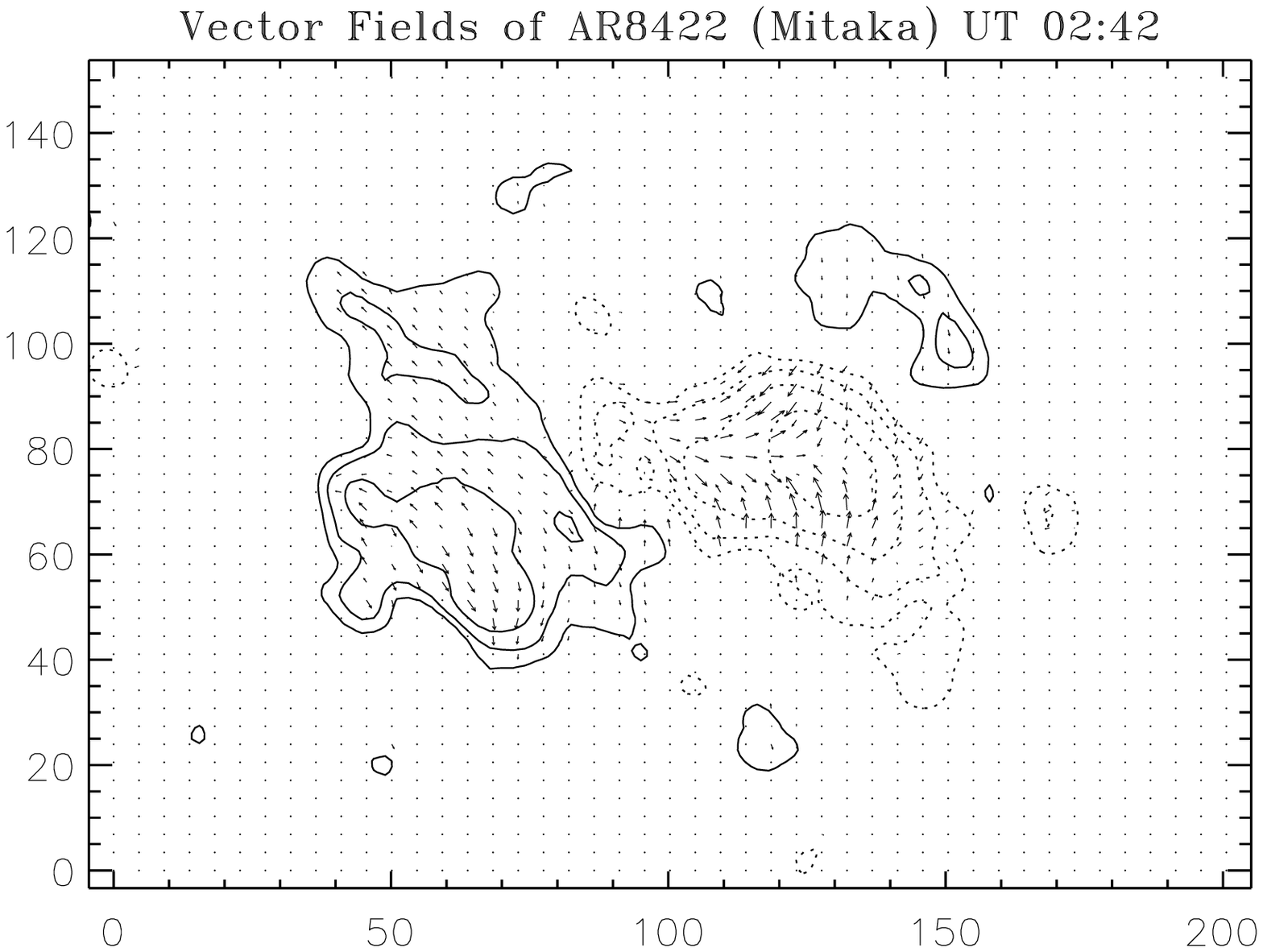,height=6cm,width=8cm}
\caption{
Comparison of BOAO(upper) and Mitaka(lower) vector magnetograms 
of AR 8422 observed on Dec. 28, 1998.
In the all figures, the  solid lines stand for  the   
positive longitudinal polarities and  the   dotted lines for 
the negative polarities. 
The contour levels correspond to 100, 200, 400, 800,
and 1600 G, respectively.    
The   length of arrows     represents   the magnitude of   transverse 
field component with $B_{tmax}=939$ G. 
}
\end{figure}  
\begin{figure}
\psfig{figure=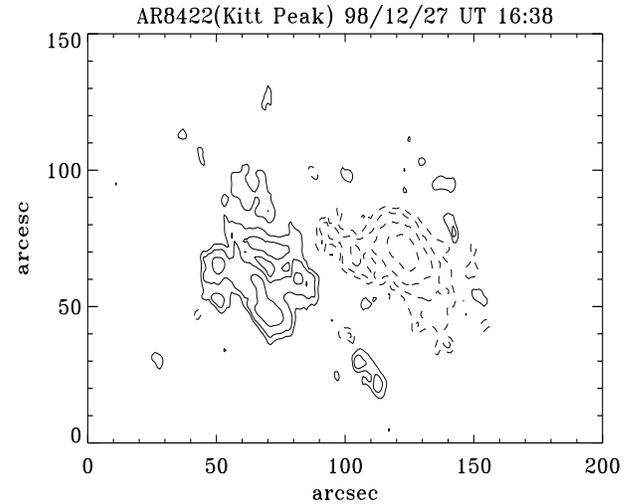,height=6.8cm,width=8.5cm}
\caption{
Longitudinal magnetogram of AR 8422 observed on Dec. 27, 1998
at Kitt Peak Solar Observatory. 
Its main features are very similar to those of SOFT's and Mitaka's
corresponding ones in Figure 2.
The contour levels and polarities of magnetogram are the same as 
those in Figure 2.
}
\end{figure}  

\begin{figure}
\psfig{figure=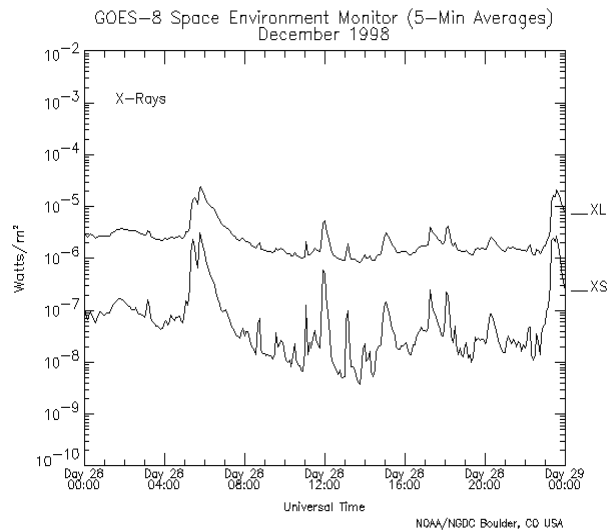,height=7cm,width=8.5cm}
\caption{
GOES X-ray fluxes of Dec. 28, 1998.
A strong peak at UT 05:48 indicates a M3.1 X-ray  
flare occurred in the AR 8419. 
Here XL represents long X-ray fluxes (1-8$\AA$) and XS, short
X-ray fluxes (0.5-4$\AA$).}
\end{figure}  

In Figure 3
we present a longitudinal magnetogram  observed on
Dec. 27, 1998  at Kitt Peak Solar
Observatory. 
Even though it was obtained about eight hours before than
the corresponding SOFT's one (Figure 2-a)), its main features are
very similar to those of SOFT's one. 
Our study shows
that our vector magnetograph should normally
work.

\begin{figure}
\psfig{figure=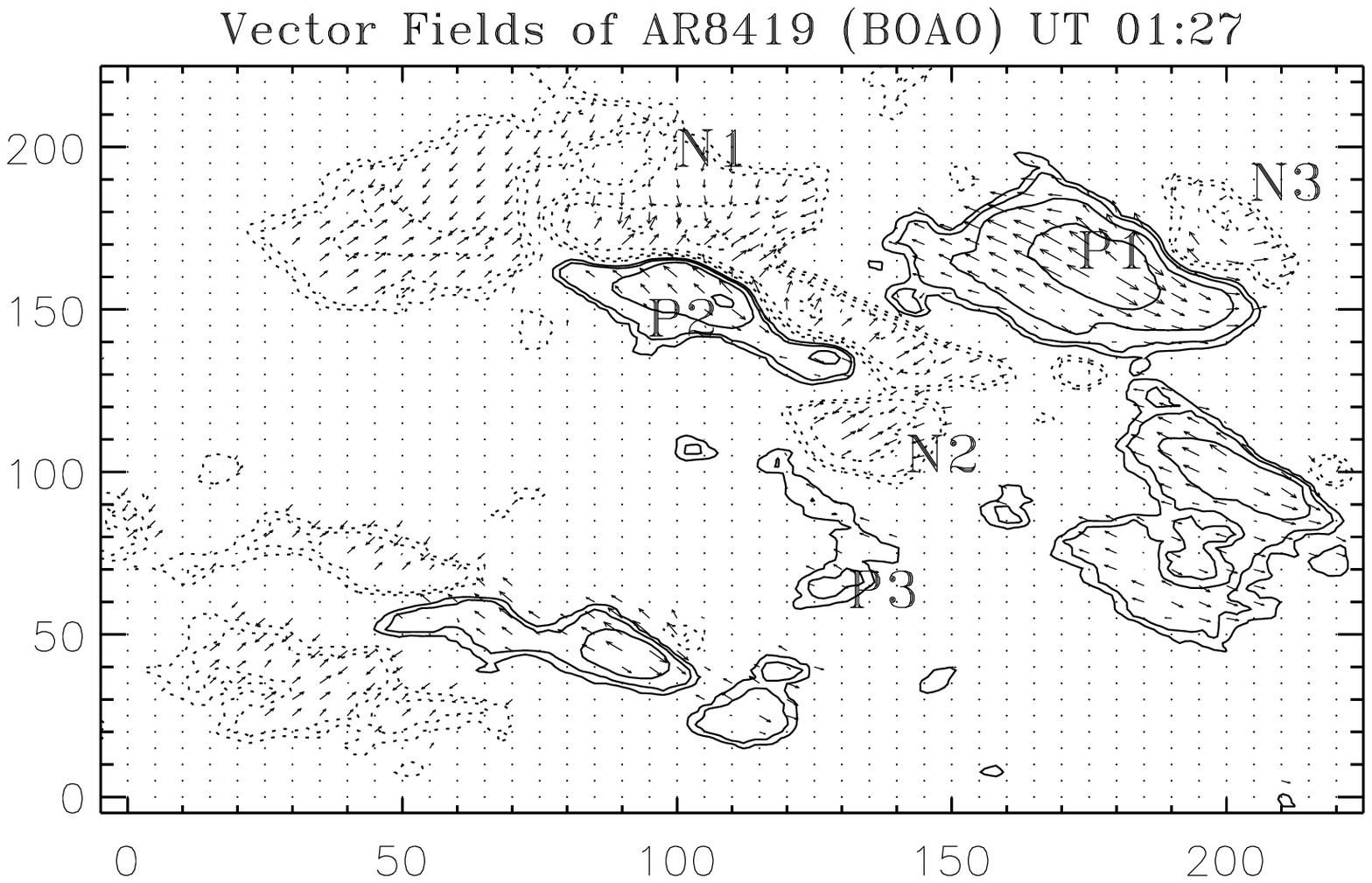,height=8cm,width=8cm}
\psfig{figure=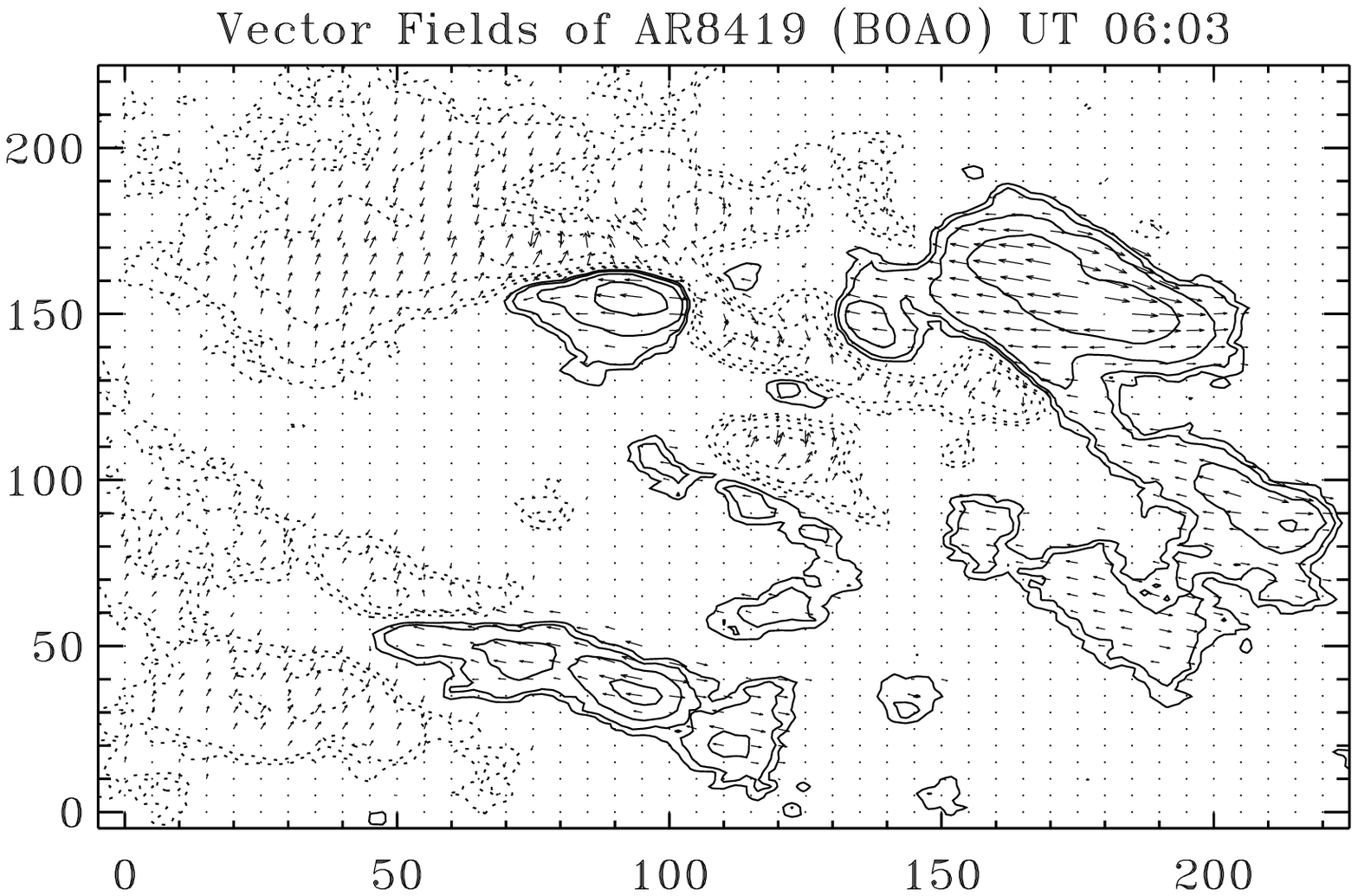,height=8cm,width=8cm}
\caption{
Vector magnetograms observed before and after a M3.1/1B
flare in AR 8419. 
The contour levels and polarities of magnetograms are the same as 
those in Figure 2.
The   length of arrows     represents   the magnitude of   transverse 
field component with $B_{tmax}=1153$ G.
}
\end{figure}  
\begin{figure}
\psfig{figure=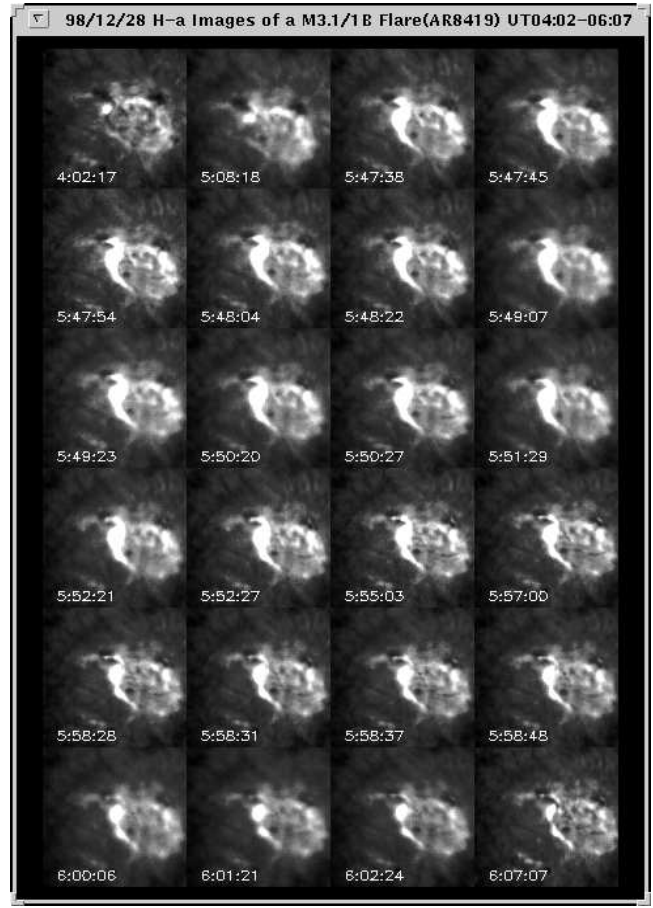,height=12cm,width=8.5cm}
\caption{
Series of BOAO H$\alpha$ filtergrams of AR 8419 during its flaring activity.
A field of view for each filtergram is $220''\times220''$.}
\end{figure}  
\begin{figure}
\psfig{figure=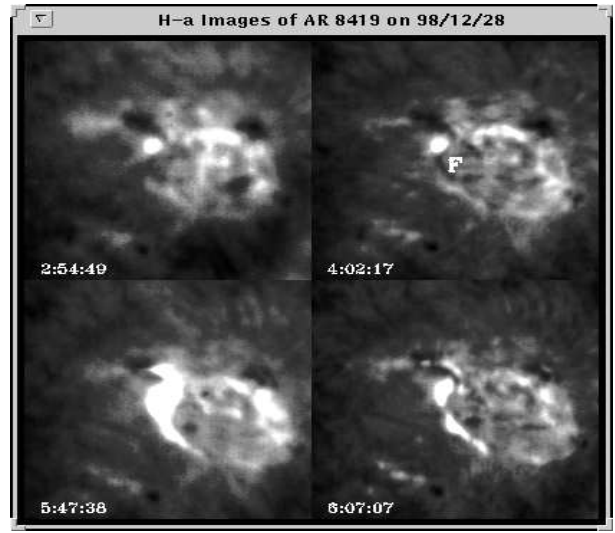,height=7cm,width=8cm}
\caption{
BOAO H$\alpha$ filtergrams of AR 8419 observed before and after
the M3.1 flare. On the image at UT 04:02, $F$ denotes a filament
eruption region, which exactly match the inversion line of the
quadrupolar configuration.}
\end{figure}

\begin{figure}
\psfig{figure=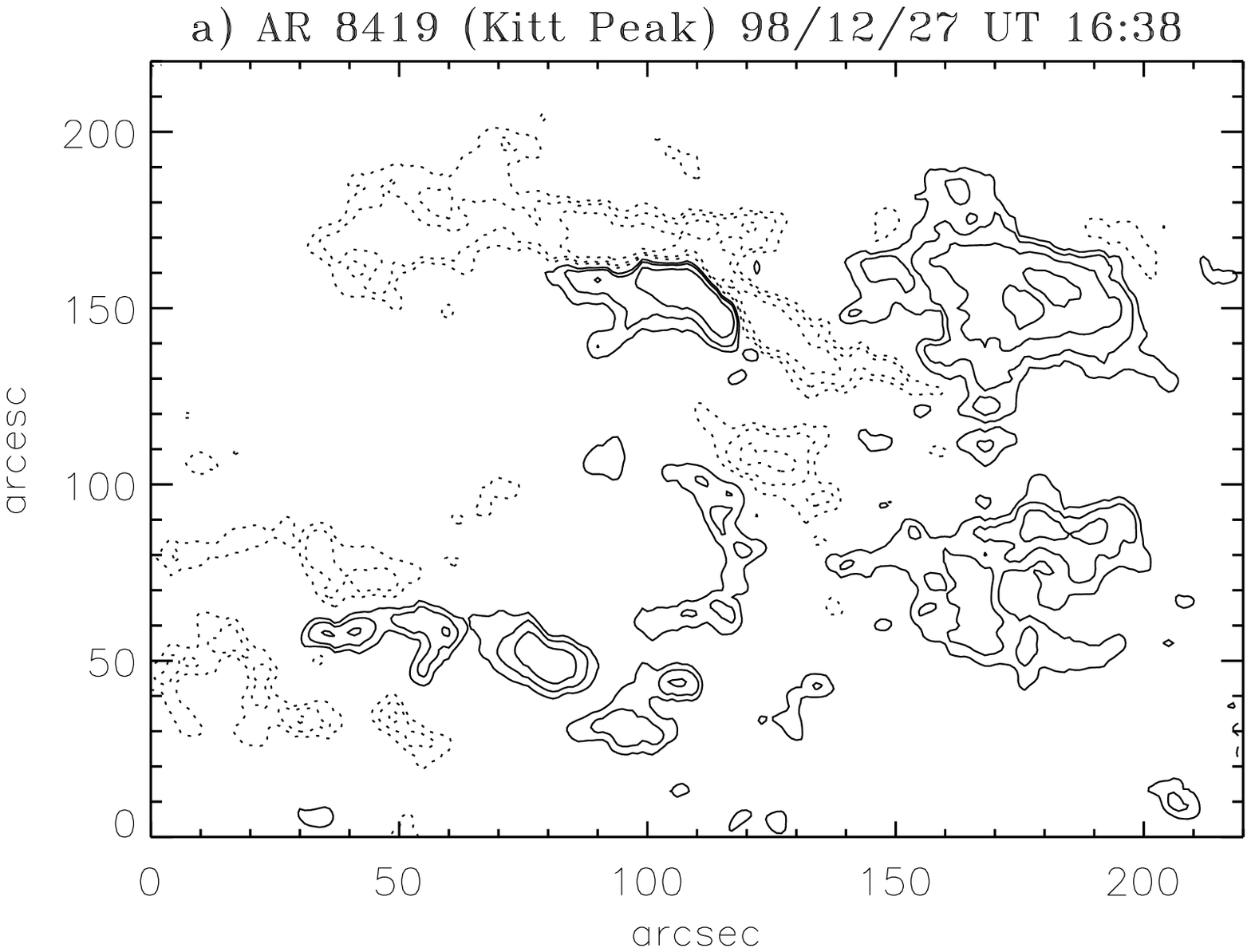,height=7cm,width=8.cm}
\psfig{figure=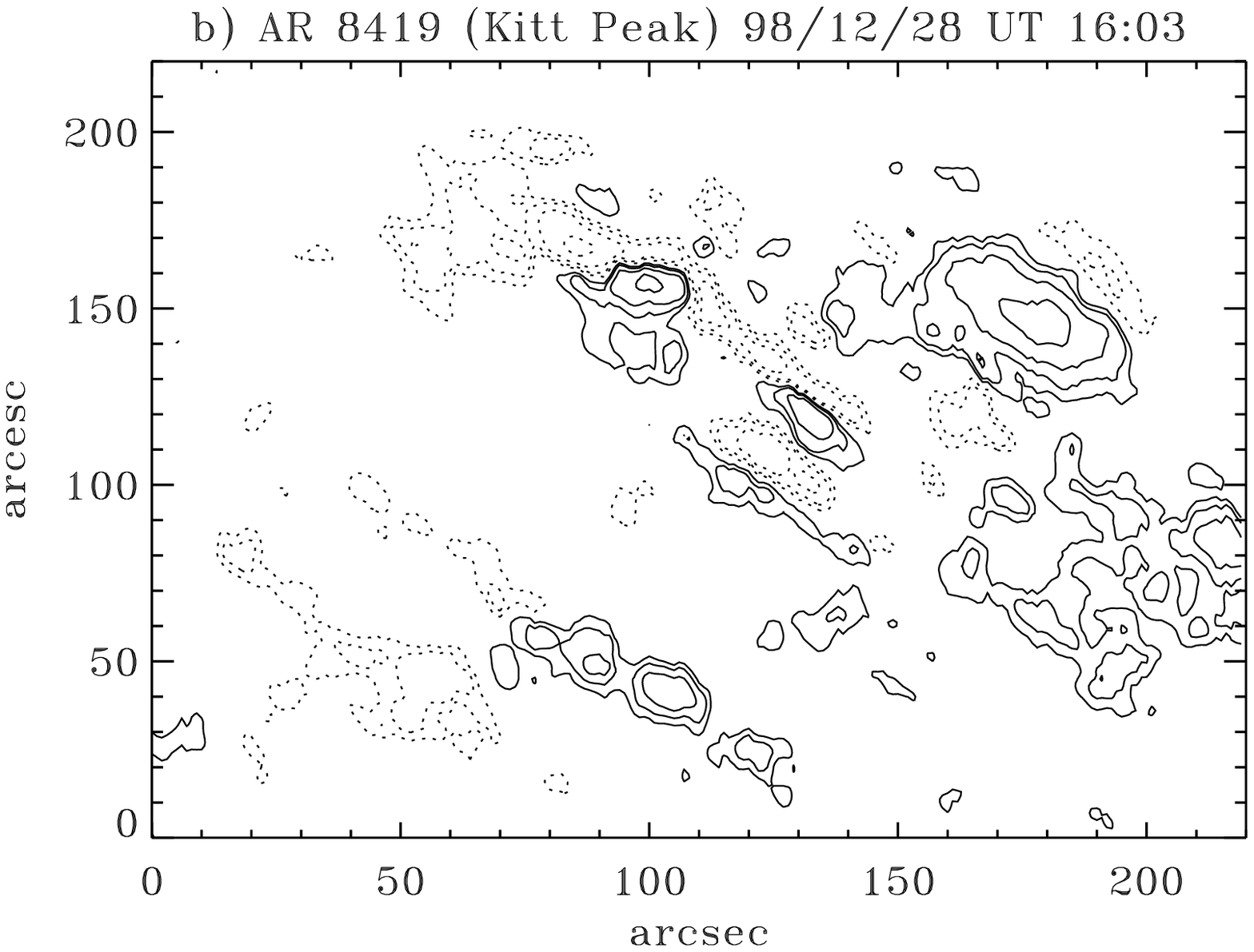,height=7cm,width=8.cm}
\caption{
Longitudinal magnetograms of AR 8419  on Dec. 27
and 28, 1998, which were made from the full disk
magnetograms of corresponding days observed at
Kitt Peak Solar Observatory.
Its main features are very similar to those of 
longitudinal fields in Figure 6. The contour levels and polarities of magnetograms are the same as 
in Figure 2.
}
\end{figure}

\section{STUDY OF ACTIVE REGION AR 8419}

We have observed vector fields together with $H\alpha$ 
and white light  of flare producing active region
AR 8419 (N27W27) 
 in which a M-class flare (M3.1/1B) occurred on Dec. 28, 1998.
According to Solar Geophysical Data, GOES X-ray flux 
 started to increase at
UT 05:45, peaked at UT 05:48, and then ended at
 UT 05:59 (See Figure 4).

At that time, the active region was of very complex magnetic polarities 
($\beta-\gamma-\delta$ type).
Figure 5 shows two vector magnetograms observed before 
and after the flare. The magnetograms were calibrated by the
second calibration method in Section II.
According to  NOAA reports,
 this region has grown in complexity as well as
 in sunspot area for three days with 
sizable proceeding (P1 in Figure 5) and following sunspot (N1).
The following sunspot has a series of
 umbra forming a NE-SW line with surrounding penumbra with a $\delta$ configuration (N1 and P2).

Figure 6  shows a time series of $H\alpha$ 
filtergrams of the SOFT for about two hours including 
the peak time of X-ray fluxes.
In the figures two dark areas correspond to the proceeding (P1) 
and following (N1) sunspots.
At UT 4:02, two small brightening patches 
were found near the $\delta$ sunspot region.
At UT 05:47, strong inverse $S$-shape (Pevtsov et al. 1997)  H$\alpha$ patches
were notified and last for
about ten minutes, and then remained several flare ribbons.

In order to examine in detail the change of H$\alpha$ 
phenomena, we present
four $H\alpha$ filtergrams observed before and after the flare in Figure 7.
It is interesting to 
note that there was a filament eruption
with inverse $S$-shape (denoted by F on the 4:02 image), 
 which exactly match
the inversion line of a quadrupolar configuration (P2, P3, N1, N2
in Figure 5). As seen on 06:07 image, several
flare ribbons are made after an eruptive phase of the flare.
It is well accepted that two flare ribbons in $H\alpha$ emission are a 
result of filament eruption along the inversion line of $\delta$ configuration region (Priest 1982, Zirin 1988, Filippov 1997).
We suggest that the M-class flare in AR 8419 should be associated with
the quadrupolar configuration and
its interaction with the new erupting filament. 

It is also found that
 sunspots in the quadrupolar configuration
were nearly in a straight line with the largely sheared inversion line
(Figure 5),
which was often observed in flare
producing active regions (Demoulin, H\'enoux,  and Mandrini 1994).
Moon et al. (1999a) showed that a large magnetic field discontinuity
exist at the separator of such a quadrupolar configuration that
directions of two bipoles are antiparallel each other ($\varphi_p=180^o$
in Table 1 of their paper).

Figure 8  shows two longitudinal magnetograms of 
AR 8419 observed on Dec. 27 and 28, 1998
at Kitt Peak Solar Observatory.
Its main features are very similar to those of 
SOFT's corresponding ones in Figure 5.
The comparison of Figure 5 and 8-a)  shows
 that a positive polarity sunspot (P2) moved westward, collide with
the following sunspot (N1) to form a $\delta$ configuration 
and to  compress longitudinal fields near the $\delta$ configuration.
In Figure 5, steep longitudinal field gradients over horizontal
direction are notified and estimated to be about 0.2 G/km 
near the $\delta$ sunspot, which was often reported in
flare-producing 
active regions (e.g., Patty and Hagyard 1986, Zhang et al. 1994 ).
It is also found that the inversion line with inverse $S$-shape become
more twisted after the flare.

\section{SUMMARY AND CONCLUSION}

We  have compared our vector fields of
AR 8422 with those made 
with a similar vector magnetograph at Mitaka 
 solar observatory.
The comparison shows that
longitudinal fields are very similar to each other
but transverse fields are a little different.
We have also compared longitudinal magnetograms
of AR 8422 and AR 8419 with Kitt Peak longitudinal
 magnetograms and confirmed that
its main features are very similar to those of the SOFT.

We have also presented
 our vector magnetograms and H$\alpha$ 
 observations of  
 AR 8419 during its flaring(M3.1/1B) activity.
Time series H-alpha observations show  a filament eruption
following the sheared inversion line of the
quadrupolar configuration of sunspots and
 an inverse $S$-shape brightening patch near the filament.
This fact imply that this flare could be associated with
the quadrupolar configuration and
 its interaction with the filament eruption.

\acknowledgements{
 We wish to thank Dr. Ichimoto and Dr. Sakurai 
 for allowing us to use some of their numerical routines
for data analysis. 
This work has been supported in part by
the Basic Research Fund (99-1-500-00 and 99-1-500-21) of 
Korea 
Astronomy Observatory and in part by the Korea-US Cooperative Science Program under KOSEF(995-0200-004-2).
}

\end{document}